\documentclass[aps,pre,twocolumn,superscriptaddress,showpacs]{revtex4}
\usepackage{amsmath}
\usepackage{graphicx}
\begin{document}
\title{Nonlinear propagation of light in Dirac matter}
\author{Bengt Eliasson}
\affiliation{Institut f\"ur Theoretische Physik,
Fakult\"at f\"ur Physik und Astronomie,
Ruhr--Universit\"at Bochum, D-44780 Bochum, Germany}
\author{P. K. Shukla}
\affiliation{RUB International Chair, International Centre for Advanced Studies in Physical Sciences,
Fakult\"at f\"ur Physik und Astronomie, Ruhr--Universit\"at Bochum, D-44780 Bochum, Germany}
\received{13 May 2011; revised 10 July 2011}
\begin{abstract}
The nonlinear interaction between intense laser light and a quantum plasma is modeled
by a collective Dirac equation coupled with the Maxwell equations. The model is used to study
the nonlinear propagation of relativistically intense laser light in a quantum plasma including the
electron spin-1/2 effect. The relativistic effects due to the high-intensity laser light lead, in general,
to a downshift of the  laser frequency,  similar to a classical plasma where the relativistic mass increase leads
to self-induced transparency of laser light and other associated effects. The electron spin-1/2 effects lead to a
frequency up- or downshift of the electromagnetic (EM) wave, depending on the spin state of the plasma and the polarization
of the EM wave.  For laboratory solid density plasmas, the spin-1/2 effects on the propagation of light
are small,  but they may be significant in super-dense plasma in the core of white dwarf stars.
We also discuss extensions of the model to include kinetic effects of a distribution of
the electrons on the nonlinear propagation of EM waves in a quantum plasma.
\end{abstract}

\maketitle

\section{Introduction}

The introduction of intense lasers has lead to a great variety of applications, including plasma based particle acceleration
to relativistic energies \cite{Bingham03,Mangles04}, and with X-ray free-electron lasers \cite{Hand09} there are new possibilities to explore
dense matter on atomic and single molecule levels. On these length scales, of the order of a few {\AA}ngstr\"om, quantum
effects play an important role in the dynamics of the electrons. Using novel laser scattering techniques, quantum dispersive effects
have been observed experimentally  both in the degenerate electron gas in metals and in warm dense matters \cite{Glenzer}.
Hence, it is expected that quantum mechanical effects must be taken into account in  intense laser-solid density plasma interaction
experiments \cite{Andreev,Bulanov,MarklundShukla}, and in  quantum free-electron laser systems \cite{Serbeto08,Serbeto09,Piovella08}.
Even though $\gamma$-ray lasers have not yet been manufactured, there have been suggestions that such lasers could be realized by means
of annihilation of Bose-Einstein condensated positronium \cite{Mills04,Cassidy07}, or by the excitation and nuclear spin relaxation in a lattice
of thorium atoms \cite{Tkalya11}. This would lead to a new regime of intense laser-plasma interactions, where the relativistic quantum dynamics
plays a decisive role. Intense x-ray and $\gamma$-ray sources exist naturally in astrophysical objects in the form of x-ray and $\gamma$-ray
repeaters, etc. \cite{Chabrier,Coe,Hurley}.  In the past, the linear plasma response for relativistic (i.e. relativistically distributed) quantum plasmas
were studied by deriving the longitudinal and transverse response functions for mildly and strongly  degenerate electron
distributions \cite{Tsytovich61,Jancovici62}. It was noted \cite{Tsytovich61} that for super-dense plasmas where $\hbar\omega_{pe}>2m_e c^2$,
there is a possibility of collisionless pair creation, where $\hbar$ is the Planck constant divided by $2\pi$, $\omega_{pe}$ the electron plasma frequency,
$m_e$ the electron mass, and $c$ the speed of light in vacuum. The results were extended by using a Wigner functions approach \cite{Hakim78a,Hakim78b,Hakim80,Sivak85},
and by  considering the longitudinal response \cite{Delsante80,Kowalenko85}, and more general results for different
distribution functions have also been obtained \cite{Hayes84,Melrose84,Melrose06}.
Relativistic quantum fluid models have recently been derived \cite{Asenjo11}, partially based on earlier works \cite{Takabayasi} of
fluid-like formulations of the Dirac equation. When the intensity of the electromagnetic (EM) wave reaches a critical level
(e.g. around $10^{19}\,\mathrm{W/cm}^2$ for one micron wavelength lasers),  the relativistic electron mass increase and the associated
nonlinearity plays a significant role for the propagation and dynamics of the EM wave \cite{Akhiezer56}. In addition, the relativistic
ponderomotive force \cite{Shukla} produces density modifications in the plasma, and the combined effects of the relativistic electron mass increase
and relativistic ponderomotive force can lead to a modulational instability and collapse localization of EM waves \cite{McKinstrie89,Tsintsadze91}.
Clearly, for intense EM waves interacting with the plasma in the X-ray and $\gamma$-ray regimes, both relativistic and quantum effects must be
taken into account on an equal footing.

In this paper, we present a nonlinear model, based on the Dirac equation coupled with the Maxwell equations that are capable of treating both the
relativistic (propagation and mass increase), quantum (tunneling/diffraction) effects, and electron spin effects.
The mathematical aspects of this system has been
studied in the past \cite{Gross66}.
We will here use the basic model to investigate
the nonlinear propagation of large amplitude EM waves in a quantum plasma  with different spin polarizations. Our work has potential applications in
laser-matter experiments \cite{Glenzer,Malkin07}, quantum free-electron laser systems  \cite{Serbeto08,Serbeto09,Piovella08}, as well as in astrophysical
environments \cite{Chabrier,Coe,Hurley}.

\section{The mathematical model}

The quantum mechanical description of the relativistic dynamics of an electron in an EM field
is given by the Dirac equation
\begin{equation}
{\cal W}\psi-c\boldsymbol{\alpha}\cdot{\cal P}\psi-m_e c^2 \beta \psi=0,
\label{Dirac}
\end{equation}
where we have defined the energy and momentum operators as
\begin{equation}
  {\cal W}=i\hbar\frac{\partial}{\partial t}+e\phi,
\end{equation}
and
\begin{equation}
  {\cal P}=-i\hbar\nabla+e{\bf A},
\end{equation}
respectively. Here, $\phi$ and ${\bf A}$ are the scalar and vector potentials, and $e$ is the magnitude of the electron charge.
The vector $\boldsymbol{\alpha}=\alpha_x\widehat{\bf x}+\alpha_y\widehat{\bf y}+\alpha_z\widehat{\bf z}$,
where $\widehat{\bf x}$, $\widehat{\bf y}$ and $\widehat{\bf z}$ are unit vectors in the $x$, $y$ and $z$ directions,
have components consisting of the Dirac matrices
\begin{equation}
  \alpha_k=\left(
  \begin{matrix}
  0 & \sigma_k
  \\
  \sigma_k & 0
  \end{matrix}
  \right), \qquad k=x,\,y,\,z,
\end{equation}
where the Pauli spin matrices are
\begin{equation}
  \sigma_x=\left(
  \begin{matrix}
  0 & 1
  \\
  1 & 0
  \end{matrix}
  \right), \quad
  \sigma_y=\left(
  \begin{matrix}
  0 & -i
  \\
  i & 0
  \end{matrix}
  \right), \quad
  \sigma_z=\left(
  \begin{matrix}
  1 & 0
  \\
  0 & -1
  \end{matrix}
  \right),
\end{equation}
and the matrix $\beta$ reads
\begin{equation}
  \beta=\left(
  \begin{matrix}
  {\sf I} & 0
  \\
  0 & -{\sf I}
  \end{matrix}
  \right),
\end{equation}
with ${\sf I}$ being unit $2\times 2$ matrices.

We now wish to use the charge and current densities as sources for the self-consistent EM scalar and vector potentials for a quantum plasma. We therefore let the 4-spinor $\psi=[\psi_1\ \psi_2\ \psi_3\ \psi_4]^{T}$ represent an ensemble of electrons ($T$ denotes the transpose of the matrix).
The electric charge and current densities are obtained as
\begin{equation}
\rho_e=-e \psi^\dagger \psi=-e\sum_{j=1}^4 |\psi_j|^2,
\label{rho_e}
\end{equation}
and
\begin{equation}
{\bf j}_e=- e \psi^\dagger c\boldsymbol{\alpha} \psi,
\label{j_e}
\end{equation}
respectively (where $\psi^\dagger=[\psi_1^\ast\ \psi_2^\ast\ \psi_3^\ast\ \psi_4^\ast]$).
The current density incorporates both the particle current and spin current. The charge and current densities obey the continuity equation
\begin{equation}
  \frac{\partial \rho_e}{\partial t}+\nabla\cdot {\bf j}_e=0.
  \label{continuity}
\end{equation}

The self-consistent vector and scalar potentials are obtained from the EM wave equations
\begin{equation}
\frac{\partial^2{\bf A}}{\partial t^2}+ c^2 \nabla\times(\nabla\times{\bf A})
+  \nabla\frac{\partial \phi}{\partial t}=\mu_0 c^2{\bf j}_e,
\label{vector_pot}
\end{equation}
and
\begin{equation}
   \nabla^2\phi+\nabla\cdot\frac{\partial {\bf A}}{\partial t}
=-\frac{1}{\varepsilon_0}(\rho_e+\rho_i),
   \label{scalar_pot}
\end{equation}
where $\mu_0$ is the magnetic vacuum permeability, $\varepsilon_0$ is the electric permittivity
in vacuum, $c=1/\sqrt{\varepsilon_0 \mu_0}$, and $\rho_i$ is
the neutralizing positive charge density of the ions. For immobile,
singly charged ions, we have $\rho_i=e n_0$, where $n_0$ is the equilibrium ion number
density. In our model, we have neglected the fact that degenerate, cold electrons are
distributed uniformly in momentum space up to the Fermi sphere. The Fermi pressure plays an important role in the dynamics of longitudinal electrostatic waves, where it contributes to the dispersion of the waves. For transverse electromagnetic waves, which will be our main interest here,
the distribution of electrons play a minor role. The Fermi pressure is unimportant since the transverse electromagnetic waves are not associated with density perturbations. The effects on the current of particles streaming in one direction is canceled by particles streaming in the opposite direction so that the net effect on the electromagnetic wave is negligible. For extremely dense plasmas, where $\hbar \omega_{pe}$ is comparable to $m_e c^2$, the speeds of the electrons on the Fermi sphere become relativistic and one can expect a frequency downshift due to the relativistic mass increase of these electrons. This effect, however, is outside the scope of our model.

\section{Circularly polarized light in Dirac matter}

We here consider the nonlinear propagation of light in Dirac matter with different spin
polarizations. Solutions have been obtained in the past \cite{Volkov35,Felber75}
for single particles in an EM field. Here we formulate the problem in
a plasma environment, where we require the quasi-neutrality and current-neutrality
along the propagation direction of the EM field.
The plasma environment introduces a dimensionless quantum parameter
\begin{equation}
  H=\frac{\hbar \omega_{pe}}{m_e c^2},
\label{H}
\end{equation}
which compares the plasmonic energy $\hbar\omega_{pe}$ to the electron rest mass  energy $m_e c^2$. Typical values
are $H=10^{-4}$ for the electron number density $n_e \sim 10^{30}\,\mathrm{m}^{-3}$ in solid density
laser-plasma experiments and $H=0.007$ may be representative of the modern laser-high density matter experiments \cite{Azechi91,Kodama01,Holmlid09}
with $n_e\sim 10^{34}\,\mathrm{m}^{-3}$.  This corresponds to $\omega_{pe}=8\times 10^{16}\,\mathrm{s^{-1}}$
and $\lambda_e=4\times 10^{-9}\,\mathrm{m}$ for $H=10^{-4}$, and $\omega_{pe}=5.4\times 10^{18}\,\mathrm{s}^{-1}$
and $\lambda_e=5.5\times 10^{-11}\,\mathrm{m}$ for $H=0.007$, where $\lambda_e=c/\omega_{pe}$ is the electron
skin depth. On the other hand, in extremely dense plasmas in the core of white dwarf stares, the
quantum parameter $H$ may be of the order unity. For $H>2$, it has been noticed \cite{Tsytovich61} that there is
a possibility of pair creation, and one then has to take into account positrons on the plasma dynamics.
We do not consider this case here, since it is beyond the scope of our model.

\subsection{Solution of the Dirac equation}

The first step to a self-consistent picture is to solve the Dirac equation for a circularly polarized EM wave with constant amplitude.
We assume a right-hand circularly polarized EM wave of the form
\begin{equation}
  {\bf A}=A_0(\widehat{\bf x}\cos \theta-\widehat{\bf y}\sin \theta),\mbox{ where } \theta=k_0 z-\omega_0 t,
  \label{CPEM}
\end{equation}
where the frequency and wavenumber $\omega_0$ and $k_0$ are constants, and we assume that $\phi=0$. In this case, the Dirac equation can be formulated
into an eigenvalue problem (See Appendix \ref{App_eig})
\begin{align}
\begin{split}
&\big[\hbar(\Omega+\frac{\omega_0}{2})-m_e c^2\big]\widetilde{\Psi}_1-
c\hbar (K +\frac{k_0}{2})\widetilde{\Psi}_3
\\
&-ceA_0 \widetilde{\Psi}_4=0,
\end{split}
\label{Dirac_exp1}
\\
\begin{split}
&\big[\hbar(\Omega-\frac{\omega_0}{2})-m_e c^2 \big]\widetilde{\Psi}_2+
c\hbar (K-\frac{k_0}{2})\widetilde{\Psi}_4
\\
&-ceA_0 \widetilde{\Psi}_3=0,
\end{split}
\label{Dirac_exp2}
\\
\begin{split}
&\big[\hbar(\Omega+\frac{\omega_0}{2})+m_e c^2\big]\widetilde{\Psi}_3-
c\hbar (K+\frac{k_0}{2})\widetilde{\Psi}_1
\\
&-ceA_0 \widetilde{\Psi}_2=0,
\end{split}
\label{Dirac_exp3}
\intertext{and}
\begin{split}
&\big[\hbar(\Omega-\frac{\omega_0}{2})+m_e c^2\big]\widetilde{\Psi}_4+
c\hbar (K-\frac{k_0}{2})\widetilde{\Psi}_2
\\
&-ceA_0 \widetilde{\Psi}_1=0,
\end{split}
\label{Dirac_exp4}
\end{align}
for the constant spinor components $\widetilde{\Psi}_1$--$\widetilde{\Psi}_4$
where $\Omega$ takes the role of an eigenvalue for given values of $K$. The coefficient matrix is real and symmetric, hence $\Omega$ is real and $\widetilde{\Psi}_j$ can be taken to be real.

Eliminating $\widetilde{\Psi}_1$--$\widetilde{\Psi}_4$ in (\ref{Dirac_exp1})--(\ref{Dirac_exp4}), we obtain the characteristic equation
\begin{equation}
D_+ D_-
+(\omega_0^2-c^2 k_0^2)\frac{c^2 e^2 A_0^2}{\hbar^2}=0,
\label{char}
\end{equation}
where
\begin{equation}
  D_\pm=\bigg(\Omega\pm\frac{\omega_0}{2}\bigg)^2-c^2\bigg(K\pm\frac{k_0}{2}\bigg)^2-\frac{m_e^2 c^4 \gamma_0^2}{\hbar^2},
\end{equation}
and we have denoted $\gamma_0=(1+{e^2 A_0^2}/{m_e^2 c^2})^{1/2}$.
We note that $D_+$ and $D_-$ are Klein-Gordon operators that are coupled in Eq.~(\ref{char}), but which become uncoupled in the vacuum case $\omega_0=c k_0$.

\subsection{The nonlinear plasma susceptibility and the dispersion relation for waves}

We here discuss the collective plasma response in the presence of the EM wave.
We require quasi-neutrality $\rho_e+\rho_i=0$ in the plasma, which leads to
\begin{equation}
  \sum_{j=1}^4|\psi_j|=\sum_{j=1}^4 \widetilde{\Psi}_j^2=n_0\equiv \Psi_0^2.
  \label{quasi_neutrality}
\end{equation}
The current density is (see Appendix \ref{App_curr})
\begin{equation}
  \begin{split}
  {\bf j}_e =&-2ec[(\widetilde{\Psi}_1\widetilde{\Psi}_4+\widetilde{\Psi}_3\widetilde{\Psi}_2)(\widehat{\bf x}\cos\theta-\widehat{\bf y}\sin\theta) \\
  &+(\widetilde{\Psi}_1\widetilde{\Psi}_3-\widetilde{\Psi}_2\widetilde{\Psi}_4)\widehat{\bf z}].
  \end{split}
\end{equation}
For our study, it is natural to require that the system is at rest in the $z$ direction, so that $j_z=0$, i.e.
\begin{equation}
  \widetilde{\Psi}_1\widetilde{\Psi}_3-\widetilde{\Psi}_2\widetilde{\Psi}_4=0.
  \label{current_neutrality}
\end{equation}
The resulting current density ${\bf j}_e=-2ec(\widetilde{\Psi}_1\widetilde{\Psi}_4+\widetilde{\Psi}_3\widetilde{\Psi}_2)(\widehat{\bf x}\cos\theta-\widehat{\bf y}\sin\theta)$ is
right-hand circularly polarized, similar to the vector potential ${\bf A}$ in (\ref{CPEM}).

Inserting the expressions for the circularly polarized current ${\bf j}_e$ and vector potential ${\bf A}$ into the EM wave equation (\ref{vector_pot}) with $\phi=0$, gives
\begin{equation}
  (\omega_0^2-c^2k_0^2)A_0=2\mu_0 e c^3(\widetilde{\Psi}_1\widetilde{\Psi}_4+\widetilde{\Psi}_3\widetilde{\Psi}_2),
  \label{w0_k0_1}
\end{equation}
which shows the dependence between $\omega_0$ and $k_0$. We can also write Eq.~(\ref{w0_k0_1}) as
\begin{equation}
  \omega_0^2-c^2k_0^2=-\chi_e\omega_0^2,
  \label{w0_k0}
\end{equation}
where
\begin{equation}
  \chi_e=-\frac{2 ec (\widetilde{\Psi}_1\widetilde{\Psi}_4+\widetilde{\Psi}_3\widetilde{\Psi}_2)}{\varepsilon_0\omega_0^2 A_0}
  \label{chi_e}
\end{equation}
is the electric susceptibility of the quantum plasma. Equation (\ref{w0_k0}), together with the quasi-neutrality and current-neutrality conditions (\ref{quasi_neutrality}) and (\ref{current_neutrality}), and the Dirac system (\ref{Dirac_exp1})--(\ref{Dirac_exp4}) forms a self-consistent system for the unknowns $\Psi_1$--$\Psi_4$, $\Omega$, $K$, and $\omega_0$ for given values of $H$, $A_0$ and $k_0$.

\subsection{Polar representation of the Dirac equation}

The quasi-neutrality condition (\ref{quasi_neutrality}) suggests that $\widetilde{\Psi}_1$--$\widetilde{\Psi}_4$ could be
represented with a polar representation. In addition, we wish the current-neutrality condition (\ref{current_neutrality}) to be fulfilled.
Both these conditions are fulfilled if we make the special choice of polar representation $\widetilde{\Psi}_j=(\sqrt{n_0}/2)Y_j$ with
\begin{align}
&Y_1=\cos \varphi_2+\sin\varphi_1
\\
&Y_2=\cos \varphi_1+\sin\varphi_2
\\
&Y_3=\cos \varphi_2-\sin\varphi_1
\\
&Y_4=\cos \varphi_1-\sin\varphi_2.
\end{align}
The electron susceptibility (\ref{chi_e}) then takes the simple form
\begin{equation}
  \chi_e=-\frac{e c n_0}{\varepsilon_0 \omega_0^2 A_0}\cos(\varphi_1+\varphi_2),
\end{equation}
and Eqs. (\ref{Dirac_exp1})--(\ref{Dirac_exp4}) become, respectively,
\begin{align}
\begin{split}
&\big[\hbar(\Omega+\frac{\omega_0}{2})-m_e c^2\big]Y_1-c\hbar (K+\frac{k_0}{2})Y_3
\\
&-ceA_0 Y_4=0
\end{split}
\label{Dirac_p1}
\\
\begin{split}
&\big[\hbar(\Omega-\frac{\omega_0}{2})-m_e c^2\big]Y_2+c\hbar (K-\frac{k_0}{2})Y_4
\\
&-ceA_0 Y_3=0
\end{split}
\label{Dirac_p2}
\\
\begin{split}
&\big[\hbar(\Omega+\frac{\omega_0}{2})+m_e c^2\big]Y_3-c\hbar (K+\frac{k_0}{2})Y_1
\\
&-ceA_0 Y_2=0
\end{split}
\label{Dirac_p3}
\intertext{and}
\begin{split}
&\big[\hbar(\Omega-\frac{\omega_0}{2})+m_e c^2\big]Y_4+c\hbar (K-\frac{k_0}{2})Y_2
\\
&-ceA_0 Y_1=0,
\end{split}
\label{Dirac_p4}
\end{align}
which, coupled with Eq. (\ref{w0_k0}), gives the unknowns $\varphi_1$, $\varphi_2$, $\Omega$, $K$, and $\omega_0$ for given values of $A_0$, $k_0$ and $H$. The general solution is difficult to find in terms of simple expressions, but can be evaluated numerically with standard methods, e.g. Newton iterations.
Some special choices of $\varphi_1$ and $\varphi_2$, and the corresponding $\widetilde{\Psi}_1$--$\widetilde{\Psi}_4$ are shown in the table below:
\begin{center}
\begin{tabular}{|c|c|c|c|c|c|}
\hline
$\varphi_1$ & $\varphi_2$ & $\widetilde{\Psi}_1$ & $\widetilde{\Psi}_2$ & $\widetilde{\Psi}_3$ & $\widetilde{\Psi}_4$
\\
\hline
$\pi/2$ & $0$ & $\sqrt{n_0}$ & $0$ & $0$ & $0$
\\
$0$ & $\pi/2$ & $0$ & $\sqrt{n_0}$ & $0$ & $0$
\\
$-\pi/2$ & $0$ & $0$ & $0$ & $\sqrt{n_0}$ & $0$
\\
$0$ & $-\pi/2$ & $0$ & $0$ & $0$ & $\sqrt{n_0}$
\\
\hline
\end{tabular}
\end{center}
The two first lines correspond to positive energy states of the two spin polarizations (spin 'up' and spin 'down'), while the two last
lines correspond to negative energy states, or pair states.

\subsection{Special cases of plasma susceptibilities}

We now consider some special cases where we can find simple expressions for the electron susceptibility of the plasma, as
well as some numerical solutions of the fully nonlinear case.

\subsubsection{Linear and nonlinear propagation of light}

We first consider linear propagation of waves where $e A_0/m_e c\ll 1$. We linearize the system by setting
$\Psi_j=\Psi_j^{(0)}+\Psi_j^{(1)}$, where $|\Psi_j^{(0)}|\gg|\Psi_j^{(1)}|$, while $A_0=A_0^{(1)}$ is a first order quantity.
For the zeroth order Dirac equation, we thus set $A_0$ to zero and $\Psi_j=\Psi_j^{(0)}$ in (\ref{Dirac_exp1})--(\ref{Dirac_exp4}).
Two possible solutions of the resulting system are found by choosing $\widetilde{\Psi}_1^{(0)}=\Psi_0$ (where $\Psi_0=\sqrt{n_0}$)
and $\widetilde{\Psi}_2^{(0)}=\widetilde{\Psi}_3^{(0)}=\widetilde{\Psi}_4^{(0)}=0$, or $\widetilde{\Psi}_2^{(0)}=\Psi_0$
and $\widetilde{\Psi}_1^{(0)}=\widetilde{\Psi}_3^{(0)}=\widetilde{\Psi}_4^{(0)}=0$, with $K=\mp k_0/2$ and $\Omega=\mp \omega_0/2+m_ec^2/\hbar$
where the upper sign corresponds to $\widetilde{\Psi}_1$ nonzero and the lower sign to $\widetilde{\Psi}_2$ nonzero.
(Other solutions also exist with $\widetilde{\Psi}_3^{(0)}$ or $\widetilde{\Psi}_4^{(0)}$ non-zero,
which correspond to pair states and which we, however, do not consider here.)
Considering the first-order quantities in  (\ref{Dirac_exp1})--(\ref{Dirac_exp4}), where we neglect first-order quantities multiplied
by each other, we find for the case with $\widetilde{\Psi}_1$ nonzero that
$\widetilde{\Psi}_3^{(1)}=0$, $-\hbar \omega_0 \widetilde{\Psi}_2^{(1)}-c\hbar k_0\widetilde{\Psi}_4^{(1)}=0$, and
$(-\hbar\omega_0+2 m_e c^2)\widetilde{\Psi}_4^{(1)}-c\hbar k_0\widetilde{\Psi}_2^{(1)}-ce A_0^{(1)} \widetilde{\Psi}_1^{(0)}=0$, from which we have
$\widetilde{\Psi}_4^{(1)}=ceA_0^{(1)}\Psi_0 \omega_0/(2\omega_0 m_e c^2-\hbar\omega_0^2+\hbar c^2 k_0^2)$, while for the case with
$\Psi_2$ nonzero, we instead have $\widetilde{\Psi}_4^{(1)}=0$, $\hbar \omega_0 \widetilde{\Psi}_1^{(1)}-c\hbar k_0\widetilde{\Psi}_3^{(1)}=0$, and
$(\hbar\omega_0+2 m_e c^2)\widetilde{\Psi}_3^{(1)}-c\hbar k_0\widetilde{\Psi}_1^{(1)}-ceA_0^{(1)} \Psi_0=0$, from which we find
$\widetilde{\Psi}_3^{(1)}=ceA_0^{(1)}\Psi_0 \omega_0/(2\omega_0 m_e c^2+\hbar\omega_0^2-\hbar c^2 k_0^2)$.
Inserting the resulting susceptibilities $\chi_e=\chi_{e\pm}$, where
$\chi_{e+}=-2 e c\widetilde{\Psi}_1^{(0)}\widetilde{\Psi}_4^{(1)}/\varepsilon_0 \omega_0^2 A_0^{(1)}$
and $\chi_{e-}=-2 e c\widetilde{\Psi}_2^{(0)}\widetilde{\Psi}_3^{(1)}/\varepsilon_0 \omega_0^2 A_0^{(1)}$, or

\begin{equation}
  \chi_{e\pm} =-\frac{\omega_{pe}^2}{\omega_0^2[1\mp\hbar(\omega_0^2-c^2 k_0^2)/2\omega_0 m_e c^2]},
  \label{linear_chi_e}
\end{equation}
into the dispersion relation (\ref{w0_k0}), we obtain
\begin{equation}
   \omega_0^2-c^2 k_0^2=\frac{\omega_{pe}^2}{1\mp\hbar(\omega_0^2-c^2 k_0^2)/2\omega_0 m_e c^2},
   \label{xx}
\end{equation}
which, after reordering of terms, takes the more transparent form
\begin{equation}
   \omega_0^2 -c^2 k_0^2-\omega_{pe}^2=\pm\frac{\hbar(\omega_0^2 -c^2 k_0^2)^2}{2 m_e c^2\omega_0}.
  \label{disp1}
\end{equation}
We see that in the classical limit $\hbar\rightarrow 0$, Eq.~(\ref{disp1}) gives the dispersion relation $\omega_0^2 -c^2 k_0^2-\omega_{pe}^2=0$
for the EM waves in a cold electron plasma, while in the zero density limit $\omega_{pe}\rightarrow 0$, Eq.~(\ref{disp1}) yields either
the vacuum EM wave dispersion relation $\omega_0^2 -c^2 k_0^2=0$, or the free particle equation of motion $\omega_0^2\pm 2 m_e c^2\omega_0/\hbar-c^2 k_0^2=0$.
The latter (including the pair branches) was also found from the electrostatic wave dispersion relation using the Klein-Gordon-Maxwell model \cite{Eliasson11}.

\begin{figure}[htb]
\includegraphics[width=8.5cm]{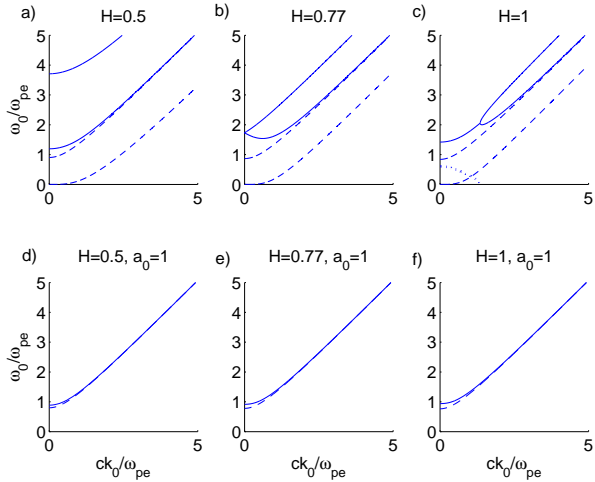}
\caption{Dispersion curves for the linear and nonlinear propagation of light in Dirac matter for
different values of $H=\hbar \omega_{pe}/m_e c^2$. For the linear cases in a)--c),
the solid and dotted curves correspond to solutions using the upper sign in Eq.~(\ref{disp1}) and dashed curves to
solutions with the lower sign in Eq.~(\ref{disp1}). Panel a) shows a high-frequency branch, the EM
branch shifted $\pm H/4$ compared to the plasma frequency at $k_0=0$, and a low-frequency branch.
For $H>4/3\sqrt{3}\approx 0.77$ the pair branch merge with the upshifted EM branch, and
there is an instability for waves with small wavenumbers; the growth rate indicated with the dotted
curve for $H=1$. Panels d)--f) shows the dispersion curves for finite amplitude ($a_0=1$) EM waves.
}
\label{Fig:linear}
\end{figure}

In Fig.~\ref{Fig:linear}a)--c), we have displayed the solutions of the linear dispersion relation (\ref{disp1}), and plotted the dispersion curves for EM waves
for different values of $H$. Figure~\ref{Fig:linear}a), for $H=0.5$, exhibits a high-frequency pair branch, the two EM
branches shifted approximately $\pm H/4$ compared to the plasma frequency at $k_0=0$, and a low-frequency branch.
For $H=H_{crit}=4/3\sqrt{3}\approx 0.77$, the pair branch merges with the upshifted EM branch for small wavenumbers, as seen in Fig.~\ref{Fig:linear}b),
and for $H>H_{crit}$ the system exhibits an instability for small wavenumbers. We have plotted the growth rate for the instability for $H=1$ in Fig.~\ref{Fig:linear}c.
Direct numerical simulations of the Dirac-Maxwell
system have confirmed this instability. It leads in the nonlinear stage to an interplay between  the $\widetilde{\Psi}_1$ and $\widetilde{\Psi}_4$ components
of the spinor, and the excitation large amplitude ($a_0=1$--$2$) oscillatory EM fields.
The dispersion relation for a finite amplitude EM wave, shown in Fig.~\ref{Fig:linear}d)--f), shows that the frequency is downshifted
in the intense EM field, and that the quantum frequency shifts decrease compared to the linear cases.
Equation (\ref{disp1}) with the lower sign also yields a low-frequency wave, plotted in
Figs.~\ref{Fig:linear}a)--\ref{Fig:linear}c) for $\omega_0^2 \ll c^2 k_0^2\ll \omega_{pe}^2$, from which we have the low-frequency dispersion relation
\begin{equation}
   \omega_0 = \frac{\hbar c^2 k_0^4}{2 m_e \omega_{pe}^2},
   \label{lowfreq}
\end{equation}
which is a low-frequency spin-EM wave. We found that the low-frequency branch exists as a propagating wave only in the weakly relativistic regime, and disappears completely as a propagating wave for $a_0\gtrsim 0.24 H$. We mention that ion dynamics can become important in the low-frequency range. We mention that ion dynamics can become important in the low-frequency range. For cold fluid ions, Eq. (\ref{w0_k0}) is replaced by $\omega_0^2-c^2k_0^2=-(\chi_e+\chi_i)\omega_0^2$ where the ion susceptibility is $\chi_i=-\omega_{pi}^2/\omega_0^2$ and $\omega_{pi}$ is the ion plasma frequency, and
$\omega_{pi}^2$ is added to the right-hand side of Eq. (\ref{xx}).
For this case, we retain (\ref{lowfreq}) for $\omega_{pi}^2$, $\omega_0^2 \ll c^2 k_0^2\ll \omega_{pe}^2$, while for $\omega_0^2 \ll c^2 k_0^2\ll \omega_{pi}^2\ll \omega_{pe}^2$, we instead have the low-frequency ion mode $\omega_0 = {\hbar k_0^2}/{2 m_i}$ where $m_i$ is the ion mass.

\subsubsection{The dipole field}

\begin{figure}[htb]
\centering
\includegraphics[width=8.5cm]{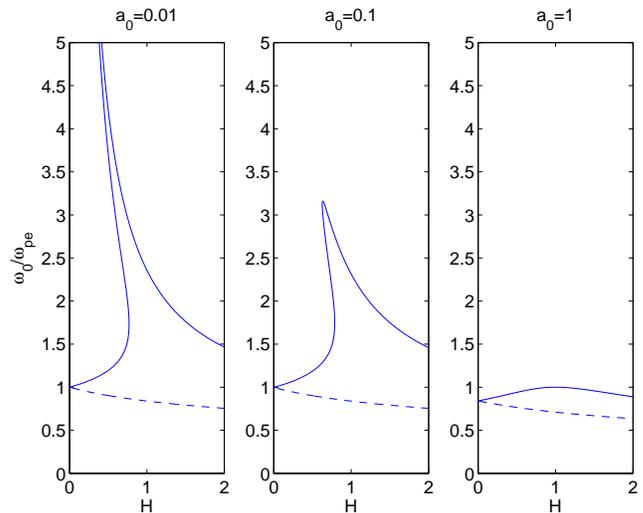}
\caption{The cutoff frequency $\omega_0$ at $K=k_0=0$ as a function of $H$ for different values
of $a_0=e A_0/m_e c$. The cutoff frequency is upshifted for the spin 'up' state (solid lines) and downshifted for spin 'down' (dashed lines).
In the classical limit $H\rightarrow 0$, we have $\omega_0\rightarrow \omega_{pe}/\sqrt{\gamma_0}$ for both spin states, where $\gamma_0=\sqrt{1+a_0^2}$.}
\label{Fig:cutoff}
\end{figure}

It is also possible to find simple expressions for the plasma susceptibility for the dipole case $k_0=K=0$ with arbitrary
amplitude $A_0$, which yields the nonlinear cutoff frequency of the EM wave. Here, inserting $\Psi_2=\Psi_3=0$ into
(\ref{Dirac_exp1})--(\ref{Dirac_exp4}) yields nontrivial solutions for
$\Omega=({1}/{\hbar})\sqrt{(m_e c^2-{\hbar\omega_{0}}/{2})^2+c^2 e^2 A_0^2}$.
Equations (\ref{Dirac_exp1}) or (\ref{Dirac_exp4}) then yield the relation between $\Psi_1$ and $\Psi_4$, which are normalized such
that $\Psi_1^2+\Psi_4^2=n_0$. On the other hand, inserting $\Psi_1=\Psi_4=0$ into
(\ref{Dirac_exp1})--(\ref{Dirac_exp4}) yields nontrivial solutions for
$\Omega=({1}/{\hbar})\sqrt{(m_e c^2+{\hbar\omega_{0}}/{2})^2+c^2 e^2 A_0^2}$.
Equations (\ref{Dirac_exp2}) or (\ref{Dirac_exp3}) then yield the relation between $\Psi_2$ and $\Psi_3$, which are normalized such
that $\Psi_2^2+\Psi_3^2=n_0$.
The resulting susceptibility is $\chi_e=\chi_{e\pm}$, where
\begin{equation}
  \chi_{e\pm}=-\frac{\omega_{pe}^2}{\omega_0^2\gamma_\pm},
\end{equation}
and we have denoted
\begin{equation}
  \gamma_\pm=\sqrt{(1\mp\frac{\hbar\omega_0}{2 m_e c^2})^2+\frac{e^2 A_0^2}{m_e^2 c^2}},
  \label{gamma}
\end{equation}
which, inserted into (\ref{w0_k0}) with $k_0=0$, yields
\begin{equation}
  \omega_0^2=\frac{\omega_{pe}^2}{\gamma_\pm}.
  \label{w0}
\end{equation}
Here $\omega_0$ can be seen  as the effective plasma frequency in the presence of quantum spin effects and the EM  field.
We note that the spin effect contributes to a relative frequency shift of the order $\pm \hbar \omega_{pe}/ 4 m_e c^2$
compared to the classical plasma frequency, while a large amplitude radiation field $A_0$ leads to a frequency
downshift, which resembles the effect of the relativistic electron mass increase in the classical plasma.

The dependence of $\omega_0$ on $H$ has been plotted in Fig.~\ref{Fig:cutoff} for different values
of $a_0=e A_0/m_e c$. We see that the frequency shifts increase linearly with $H$ for $H \ll 1$, while the
upshifted branch experiences a sharp rise at $H\approx 0.77$, which is the critical value of $H$ where the
upper branch looses its stability according to Fig.~\ref{Fig:linear}. We note that the relative quantum shift disappears both
in the classical limit $\hbar\rightarrow 0$ and in the non-relativistic limit $c\rightarrow \infty$, hence the relative shift
$\hbar \omega_{pe}/ 4 m_e c^2$ is a combined quantum and relativistic collective effect.

\subsubsection{Vacuum case}

It is also interesting to consider the vacuum case $\omega_0=c k_0$, originally considered by Volkov \cite{Volkov35},
where we have either $D_+=0$ or $D_-=0$ in Eq. (\ref{char}). We consider here ${\bf A}$ as an external field, not influenced
by the plasma, and calculate the plasma response.  For $D_+=0$, the quasi-neutrality and current-neutrality conditions (\ref{quasi_neutrality})
and (\ref{current_neutrality}) give $K=-{k_0}/{2}=-{\omega_0}/{2c}$ and $\Omega=-{\omega_0}/{2}+{\gamma_0 m_e c^2}/{\hbar}$,
and the solutions
 $\widetilde{\Psi}_1=[{(1+\gamma_0)}/{2\gamma_0}]\widetilde{\Psi}_0$,
 $\widetilde{\Psi}_2=-({e A_0}/{m_e c}){\widetilde{\Psi}_0}/{2\gamma_0}$,
$\widetilde{\Psi}_3=-({e^2 A_0^2}/{m_e^2 c^2}){\widetilde{\Psi}_0}/{2\gamma_0(1+\gamma_0)}$,
and
 $\widetilde{\Psi}_4=({e A_0}/{m_e c}){\widetilde{\Psi}_0}/{2\gamma_0}$.
If we instead use $D_-=0$, then we obtain  $K={k_0}/{2}={\omega_0}/{2c}$,
and $\Omega={\omega_0}/{2}+{\gamma_0 m_e c^2}/{\hbar}$,
and the solutions
 $\widetilde{\Psi}_1=({e A_0}/{m_e c}){\widetilde{\Psi}_0}/{2\gamma_0}$,
 $\widetilde{\Psi}_2=[{(1+\gamma_0)}/{2\gamma_0}]\widetilde{\Psi}_0$,
 $\widetilde{\Psi}_3=({e A_0}/{m_e c}){\widetilde{\Psi}_0}/{2\gamma_0}$, and
 $\widetilde{\Psi}_4=({e^2 A_0^2}/{m_e^2 c^2}){\widetilde{\Psi}_0}/{2\gamma_0(1+\gamma_0)}$.

The resulting susceptibility for both cases is
\begin{equation}
  {\chi}_e=-\frac{\omega_{pe}^2}{\omega_0^2\gamma_0},
  \label{chi_e0}
\end{equation}
which is identical to the case of the classical plasma where the relativistic gamma factor gives rise to nonlinear effects,
such as the self-induced transparency of the EM waves \cite{Akhiezer56}. The above result was also obtained in a simplified
model \cite{Eliasson11} by using the Klein-Gordon equation for spinless particles. Hence, for the vacuum case, there is no
difference in plasma response between the two spin states.

\section{Extensions to mixed states and kinetic models}

In the above investigation, we have considered the idealized case where all electrons have a well-defined spin state. Therefore, these results can be seen as limiting cases
of more complicated cases with an admixture of electrons with different spin. The simplest mixed state
solution could be achieved by assuming that we have an admixture of the electrons with spin-up and spin-down states.
If the electrons are distributed equally among the two spin states (but not among negative energy states), we
would instead of (\ref{w0_k0}) have the dispersion relation
\begin{equation}
  \omega_0^2-c^2k_0^2=-\frac{(\chi_{e+}+\chi_{e-})}{2}\omega_0^2,
  \label{w0_k0_2}
\end{equation}
where $\chi_{e+}$ and $\chi_{e-}$ are the electric susceptibilities obtained by solving the two separate Dirac equations for the two spin states, each normalized
according to (\ref{quasi_neutrality}) and fulfilling (\ref{current_neutrality}). For example, for the linear case we would use the susceptibilities $\chi_{e\pm}$ in (\ref{linear_chi_e})
in (\ref{w0_k0_2}) to construct the dispersion relation
\begin{equation}
  \omega_0^2-c^2k_0^2=-\frac{4\omega_0^2 \omega_{pe}^2 m_e^2 c^4}{\hbar^2(\omega_0^2-c^2 k_0^2)^2-4\omega_0^2 m_e^2 c^4},
\end{equation}
which can be rewritten as
\begin{equation}
  \omega_0^2-c^2 k_0^2 -\omega_{pe}^2=\frac{\hbar^2(\omega_0^2-c^2k_0^2)^3}{4\omega_0^2m_e^2 c^4}.
\end{equation}
For $\omega_0\approx\omega_{pe}$ and $ck_0\ll\omega_{pe}$, there is a relative quantum upshift of the frequency of the order
$H^2/8$, which is extremely small for normal laboratory conditions where $H\ll 1$. On the other hand, for $\omega_0^2\ll c^2 k_0^2\ll\omega_{pe}^2$,
we have a low-frequency branch $\omega_0=\hbar c^3 k_0^3/2\omega_{pe} m_e c^2$. Using a similar procedure for the nonlinear cutoff frequency, the dispersion
relation (\ref{w0}) would then be replaced by
\begin{equation}
  \omega_0^2-\frac{\omega_{pe}^2}{2}\bigg(\frac{1}{\gamma_+}+\frac{1}{\gamma_-}\bigg)=0,
\end{equation}
where $\gamma_\pm$ are given by (\ref{gamma}). More complex, kinetic models can be constructed by
extending the dispersion relation (\ref{w0_k0_2}) to include not only the sum over the two spin states, but
also sums (or integrals) over the excited states having different wavenumbers $K$. We like to mention that a kinetic model recently has been derived \cite{Mendonca11} to
 describe wave propagation in a relativistic quantum plasma based on the Klein-Gordon equation and using a Wigner transform technique.

\section{Summary and conclusions}

In this paper, we have studied the nonlinear propagation of  light in dense matter by using
a collective  Dirac model coupled with the Maxwell equations, which includes the nonlinear effects of the
finite amplitude EM waves and the electron spin-1/2 effects. As an example,
we have considered the nonlinear propagation of circularly polarized EM waves in a quantum plasma,
and have studied the effects of different spin polarization, which introduces an
up- or down shift of the EM waves, depending on whether  the plasma electron is in a spin 'up' or spin 'down' state.
This relative frequency shift is of the order $10^{-5}$--$10^{-4}$ for typical
solid density or compressed density plasmas in the laboratory, but could be
much larger in astrophysical settings (e.g. in the core of white dwarf stars), where the
plasmonic energy density $\hbar \omega_{pe}$ is comparable to the electron rest energy.
The spin related frequency shift could potentially be observed experimentally if a high-density plasma slab \cite{Holmlid09} with a definite spin state is irradiated with a laser beam with different polarization. In such a plasma it is also expected that linearly polarized laser light would perform Faraday rotation due to the different dispersive properties of the right-hand and left-hand polarized wave. For laser waves with frequencies between the cutoff frequencies for the right-hand and left-hand polarized wave, the plasma slab would work as a filter and only allow one of the polarizations to propagate through the slab.
Above a critical plasma number density, there is a density driven instability, in which the
pure spin up or spin down state is unstable under the generation of circularly polarized
EM waves. Instabilities of this type could be important in the core
of white dwarf stars, if the plasma has been spin polarized by a strong magnetic field.


\begin{widetext}
\appendix

\section{The Dirac equation for circularly polarized EM waves\label{App_eig}}
For the case of circularly polarized EM waves of the form
${\bf A}=A_0(\widehat{\bf x}\cos \theta-\widehat{\bf y}\sin \theta)$ where
$\theta=k_0 z-\omega_0 t$, we seek solutions of the Dirac equation (\ref{Dirac})
of the form $\psi=\Psi(\theta)\exp(iK z -i \Omega t)$, where we introduced the
wavenumber $K$ and frequency $\Omega$ that are related to the momentum and energy of the electrons. This yields the relations
\begin{equation}
  \frac{\partial \psi}{\partial t}=-\bigg(\omega_0\frac{d\Psi}{d\theta}+i\Omega\Psi\bigg)\exp(iKz-i\Omega t),
\end{equation}
and
\begin{equation}
  \nabla\psi=\widehat{\bf z}\bigg(k_0\frac{d\Psi}{d\theta}+i K \Psi\bigg)\exp(iK z -i \Omega t).
\end{equation}
The Dirac equation (\ref{Dirac}) takes the form
\begin{equation}
  -i\hbar(\omega_0\frac{d\Psi}{d\theta}+i\Omega\Psi)-c\boldsymbol{\alpha}\cdot\bigg[-i\hbar\widehat{\bf z}\bigg(k_0\frac{d\Psi}{d\theta}+i K \Psi\bigg)
  + e A_0(\widehat{\bf x}\cos \theta-\widehat{\bf y}\sin \theta) \Psi\bigg]-m_e c^2 \beta \Psi=0,
\end{equation}
which is a coupled system of 4 ordinary differential equations for $\Psi_1$--$\Psi_4$. To put it in an explicit form, we evaluate
\begin{equation}
  \beta \Psi=\left(
  \begin{matrix}
  1 & 0 & 0 & 0
  \\
  0 & 1 & 0 & 0
  \\
  0 & 0 & -1 & 0
  \\
  0 & 0 & 0 & -1
  \end{matrix}
  \right)
  \left(
  \begin{matrix}
  \Psi_1
  \\
  \Psi_2
  \\
  \Psi_3
  \\
  \Psi_4
  \end{matrix}
  \right)
  =
  \left(
  \begin{matrix}
  \Psi_1
  \\
  \Psi_2
  \\
  -\Psi_3
  \\
  -\Psi_4
  \end{matrix}
  \right)
\end{equation}
\begin{equation}
  \boldsymbol{\alpha}\cdot\widehat{\bf z}\Psi=\alpha_z\Psi=
\left(
  \begin{matrix}
  0 & 0 & 1 & 0
  \\
  0 & 0 & 0 & -1
  \\
  1 & 0 & 0 & 0
  \\
  0 & -1 & 0 & 0
  \end{matrix}
  \right)
  \left(
  \begin{matrix}
  \Psi_1
  \\
  \Psi_2
  \\
  \Psi_3
  \\
  \Psi_4
  \end{matrix}
  \right)
  =
  \left(
  \begin{matrix}
  \Psi_3
  \\
  -\Psi_4
  \\
  \Psi_1
  \\
  -\Psi_2
  \end{matrix}
  \right)
\end{equation}
and
\begin{equation}
  \begin{split}
  &\boldsymbol{\alpha}\cdot (\widehat{\bf x}\cos \theta-\widehat{\bf y}\sin \theta)\Psi=(\alpha_x\cos \theta-\alpha_y\sin \theta)\Psi
  \\
  &=\left[\left(
  \begin{matrix}
  0 & 0 & 0 & 1
  \\
  0 & 0 & 1 & 0
  \\
  0 & 1 & 0 & 0
  \\
  1 & 0 & 0 & 0
  \end{matrix}
  \right)\cos\theta-
\left(
  \begin{matrix}
  0 & 0 & 0 & -i
  \\
  0 & 0 & i & 0
  \\
  0 & -i & 0 & 0
  \\
  i & 0 & 0 & 0
  \end{matrix}
  \right)\sin\theta\right]
  \left(
  \begin{matrix}
  \Psi_1
  \\
  \Psi_2
  \\
  \Psi_3
  \\
  \Psi_4
  \end{matrix}
  \right)
  =
 \left(
  \begin{matrix}
  e^{i\theta}\Psi_4
  \\
  e^{-i\theta}\Psi_3
  \\
  e^{i\theta}\Psi_2
  \\
  e^{-i\theta}\Psi_1
  \end{matrix}
  \right).
  \end{split}
\end{equation}

The Dirac equation thus takes the form
\begin{align}
&\big[-i \hbar(\omega_0 \frac{d}{d \theta}+i\Omega)-m_e c^2\big]\Psi_1+ ic\hbar (k_0 \frac{d}{d\theta}+i K)\Psi_3-ceA_0 e^{i\theta}\Psi_4=0
\\
&\big[-i \hbar(\omega_0 \frac{d}{d \theta}+i\Omega)-m_e c^2\big]\Psi_2- ic\hbar (k_0 \frac{d}{d\theta}+i K)\Psi_4-ceA_0 e^{-i\theta}\Psi_3=0
\\
&\big[-i \hbar(\omega_0 \frac{d}{d \theta}+i\Omega)+m_e c^2\big]\Psi_3+ ic\hbar (k_0 \frac{d}{d\theta}+i K)\Psi_1-ceA_0 e^{i\theta}\Psi_2=0
\\
&\big[-i \hbar(\omega_0 \frac{d}{d \theta}+i\Omega)+m_e c^2\big]\Psi_4- ic\hbar (k_0 \frac{d}{d\theta}+i K)\Psi_2-ceA_0 e^{-i\theta}\Psi_1=0
\end{align}
To eliminate the $e^{i\theta}$ and $e^{-i\theta}$ phase factors, we assume $\Psi_1=\widetilde{\Psi}_1 e^{i\theta/2}$,
 $\Psi_2=\widetilde{\Psi}_2 e^{- i\theta/2}$,  $\Psi_3=\widetilde{\Psi}_3 e^{i\theta/2}$, and  $\Psi_4=\widetilde{\Psi}_4 e^{-i\theta/2}$,
where $\widetilde{\Psi}_1$--$\widetilde{\Psi}_4$ are constants, to obtain
\begin{align}
&\big[\hbar(\Omega+\frac{\omega_0}{2})-m_e c^2\big]\widetilde{\Psi}_1-
c\hbar (K +\frac{k_0}{2})\widetilde{\Psi}_3-ceA_0 \widetilde{\Psi}_4=0
\\
&\big[\hbar(\Omega-\frac{\omega_0}{2})-m_e c^2 \big]\widetilde{\Psi}_2+
c\hbar (K-\frac{k_0}{2})\widetilde{\Psi}_4-ceA_0 \widetilde{\Psi}_3=0
\\
&\big[\hbar(\Omega+\frac{\omega_0}{2})+m_e c^2\big]\widetilde{\Psi}_3-
c\hbar (K+\frac{k_0}{2})\widetilde{\Psi}_1-ceA_0 \widetilde{\Psi}_2=0
\\
&\big[\hbar(\Omega-\frac{\omega_0}{2})+m_e c^2\big]\widetilde{\Psi}_4+
c\hbar (K-\frac{k_0}{2})\widetilde{\Psi}_2-ceA_0 \widetilde{\Psi}_1=0,
\end{align}
where $\Omega$ takes the role of an eigenvalue.
The coefficient matrix is real and symmetric, hence $\Omega$ is real and $\widetilde{\Psi}_j$ can also be assumed real.
\section{Derivation of the electron current\label{App_curr}}

The $x$, $y$ and $z$ components of the current density
\begin{equation}
  {\bf j}_e = -e c \psi^\dagger\boldsymbol{\alpha}\psi
  =-e c(\psi^\dagger\alpha_x\psi\widehat{\bf x}+\psi^\dagger\alpha_y\psi\widehat{\bf y}
 + \psi^\dagger\alpha_z\psi\widehat{\bf z}),
\end{equation}
are obtained with the help of the expressions (See Appendix \ref{App_eig} for the relation between $\psi_j$ and $\widetilde{\Psi}_j$)
\begin{equation}
  \begin{split}
  \psi^\dagger\alpha_x\psi &=
\left(
  \begin{matrix}
  \psi_1^\ast
  &
  \psi_2^\ast
  &
  \psi_3^\ast
  &
  \psi_4^\ast
  \end{matrix}
  \right)
  \left(
  \begin{matrix}
  0 & 0 & 0 & 1
  \\
  0 & 0 & 1 & 0
  \\
  0 & 1 & 0 & 0
  \\
  1 & 0 & 0 & 0
  \end{matrix}
  \right)
  \left(
  \begin{matrix}
  \psi_1
  \\
  \psi_2
  \\
  \psi_3
  \\
  \psi_4
  \end{matrix}
  \right)=\psi_1^\ast\psi_4+\psi_2^\ast\psi_3+\psi_3^\ast\psi_2+\psi_4^\ast\psi_1
  \\
  &= (\widetilde{\Psi}_1\widetilde{\Psi}_4+\widetilde{\Psi}_3\widetilde{\Psi}_2)(e^{i\theta}+e^{-i\theta})
   = 2 (\widetilde{\Psi}_1\widetilde{\Psi}_4+\widetilde{\Psi}_3\widetilde{\Psi}_2) \cos(\theta),
  \end{split}
\end{equation}
\begin{equation}
  \begin{split}
  \psi^\dagger\alpha_y\psi&=
\left(
  \begin{matrix}
  \psi_1^\ast
  &
  \psi_2^\ast
  &
  \psi_3^\ast
  &
  \psi_4^\ast
  \end{matrix}
  \right)
  \left(
  \begin{matrix}
  0 & 0 & 0 & -i
  \\
  0 & 0 & i & 0
  \\
  0 & -i & 0 & 0
  \\
  i & 0 & 0 & 0
  \end{matrix}
  \right)
  \left(
  \begin{matrix}
  \psi_1
  \\
  \psi_2
  \\
  \psi_3
  \\
  \psi_4
  \end{matrix}
  \right)=-i\psi_1^\ast\psi_4+i\psi_2^\ast\psi_3-i\psi_3^\ast\psi_2+i\psi_4^\ast\psi_1
  \\
  &= i(\widetilde{\Psi}_1\widetilde{\Psi}_4+\widetilde{\Psi}_3\widetilde{\Psi}_2) (e^{i\theta}-e^{-i\theta})
   = -2 (\widetilde{\Psi}_1\widetilde{\Psi}_4+\widetilde{\Psi}_3\widetilde{\Psi}_2)\sin(\theta),
  \end{split}
\end{equation}
and
\begin{equation}
  \begin{split}
  \psi^\dagger\alpha_z\psi &=
\left(
  \begin{matrix}
  \psi_1^\ast
  &
  \psi_2^\ast
  &
  \psi_3^\ast
  &
  \psi_4^\ast
  \end{matrix}
  \right)
  \left(
  \begin{matrix}
  0 & 0 & 1 & 0
  \\
  0 & 0 & 0 & -1
  \\
  1 & 0 & 0 & 0
  \\
  0 & -1 & 0 & 0
  \end{matrix}
  \right)
  \left(
  \begin{matrix}
  \psi_1
  \\
  \psi_2
  \\
  \psi_3
  \\
  \psi_4
  \end{matrix}
  \right)=\psi_1^\ast\psi_3-\psi_2^\ast\psi_4+\psi_3^\ast\psi_1-\psi_4^\ast\psi_2
\\
  &=2 (\widetilde{\Psi}_1\widetilde{\Psi}_3-\widetilde{\Psi}_2\widetilde{\Psi}_4),
  \end{split}
\end{equation}
respectively, giving
\begin{equation}
  {\bf j}_e =-2ec[(\widetilde{\Psi}_1\widetilde{\Psi}_4+\widetilde{\Psi}_3\widetilde{\Psi}_2)(\widehat{\bf x}\cos\theta-\widehat{\bf y}\sin\theta)+(\widetilde{\Psi}_1\widetilde{\Psi}_3-\widetilde{\Psi}_2\widetilde{\Psi}_4)\widehat{\bf z}].
\end{equation}
\end{widetext}

\end{document}